\documentclass[referee]{raa}            

\usepackage{graphicx,times}             
\usepackage{natbib}
\usepackage{appendix}
\usepackage{tikz}
\usetikzlibrary{positioning, shapes.geometric}
\usepackage{amssymb,amsmath,pifont}
\usepackage{multirow}
\bibpunct{(}{)}{;}{a}{}{,}

\usepackage[a4paper=true,pagebackref=true]{hyperref}

\begin{document}

   \title{GASV: A New VLBI analysis software for Geodesy and Astrometry}

   \volnopage{Vol.0 (20xx) No.0, 000--000}      
   \setcounter{page}{1}          

   \author{Dang Yao
      \inst{1,2}
   \and Yuan-Wei Wu
      \inst{1,2}
   \and Xu-Hai Yang
      \inst{1,2}
   }


   \institute{National Time Service Center, Chinese Academy of Sciences, Xi'an, 710600, China; {\it yaodang@ntsc.ac.cn}\\
        \and
             Key Laboratory of Time Reference and Applications, Chinese Academy of Sciences,
             Xi'an, 710600, China\\
    \vs\no
   {\small Received~~20xx month day; accepted~~20xx~~month day}}

    \abstract{We present GASV, a novel Python-based software package specifically designed for the analysis of Very Long Baseline Interferometry (VLBI) data. Developed with ease of installation and user-friendliness in mind, GASV supports both pipeline and interactive processing modes. The software processes VLBI baseline delays and rates in standard formats—such as HOPS outputs and NGS card files—to estimate key geodetic and astrometric parameters, including station coordinates, Earth Orientation Parameters, source coordinates, clock parameters, and atmospheric models. We evaluate the capabilities and performance of GASV, demonstrating that its parameter estimation accuracy for IVS INT, Regular, and CONT sessions is comparable to that achieved by the VLBI analysis centers at BKG and USNO. As a state-of-the-art tool, GASV not only enables high-quality single-session data processing but also but also supports global analyses of long-term SINEX files, generating Celestial Reference Frame and Terrestrial Reference Frame solutions with reliable accuracy.
    }

   \authorrunning{D. Yao, Y.-W. Wu et al.}            
   \titlerunning{GASV}  

   \maketitle
   
   Editorial Notes: The main content of this MS is to introduce the VLBI data analysis software GASV developed by the authors, including description of the software's functions, rough structure and algorithm, and performance evaluation of the software, etc. It can be regarded as a technical report or memorandum of this software rather than an original scientific research paper. Although it is only a preliminary version and needs further improvements, it is still encouraged, especially for young researchers, to develop, publish and share such software.
   
    \section{Introduction}
    The Very Long Baseline Interferometry (VLBI) technique was originally developed by astronomers in the 1960s for imaging celestial radio sources. After over 70 years of advancement, this technique is now widely used not only in astronomy but also in geodesy, satellite tracking and deep space exploration. Unlike other space geodetic techniques-such as Satellite Laser Ranging (SLR), the Global Navigation Satellite System (GNSS), and the Doppler Orbitography and Radiopositioning Integrated by Satellite (DORIS)-all of which use artificial satellites as targets, VLBI uniquely uses extragalactic sources. This distinct  feature makes it the only technique capable of maintaining the International Celestial Reference Frame (ICRF) at radio wavelengths and measuring the complete set of Earth Orientation Parameters (EOP), which are essential for coordinate transformations between the ICRF and the International Terrestrial Reference Frame (ITRF). 
    
    EOP have a wide range of applications in astronomy (e.g., astrometry and astronomical instrument orientation), geodesy (e.g., positioning and navigation on the Earth's surface and in space), and deep space mission operations (e.g., Lunar and Mars exploration). The National Time Service Center (NTSC) of the Chinese Academy of Sciences currently operates a domestic VLBI network consisting of three 13-meter antennas, primarily used for universal time (UT1) measurement and related services \citep{Yao2020, Wu2022}. To process VLBI data effectively, we have developed software and data analysis tools for VLBI parameter determination, EOP combination, and EOP prediction. This paper presents a new VLBI analysis software, named GASV (Geodetic and Astrometric Software for VLBI).
    
    The development of geodetic VLBI data analysis tools and software dates back to the early 1970s. Based on the algorithms of \cite{Hinteregger1972} and \cite{Robertson1975}, the Goddard Space Flight Center (GSFC) developed the Calc/Solve software, written in Fortran, which has since been adopted by many VLBI analysis centers. Table \ref{analysis_software} lists the primary VLBI software packages used worldwide and their characteristics. NuSolve, a component of CALC/SOLVE, is used for producing the new vgosDB data format \citep{Bolotin2017} and performing preliminary VLBI session data analysis, but does not support global solutions. The Potsdam Open Source Radio Interferometry Tool (PORT) from the German Research Centre for Geosciences (GFZ) and the Vienna VLBI and Satellite Software (VieVS) are both coded in MATLAB and can also process satellite data observed via VLBI. VTD/Psolve\footnote{https://astrogeo.org/psolve/}, developed by Leonid Petrov at NASA's Goddard Space Flight Center, supports both interactive and batch processing of astrometric and geodetic VLBI data. Since 2022, it has been integrated into the Space Geodesy Data Analysis Software Suite (SGDASS). The Where software from the Norwegian Mapping Authority (NMA) uses Python as the interface to the SOFA\footnote{https://iausofa.org/} and the International Earth Rotation and Reference Systems Service (IERS) libraries\footnote{https://iers-conventions.obspm.fr/content/}. In contrast, GASV is entirely coded in Python and can process both single-epoch and multi-epoch data for global solutions.
    
    In this paper, we present the GASV software to the geodesy and astronomy community. Section \ref{sec2} describes the structure and algorithms of GASV. Section \ref{sec3} evaluates its performance by comparing results for INT, Regular, and CONT mode IVS data with those from established IVS data analysis centers, and also presents global solution results. Finally, Section \ref{sec4} provides conclusion and future prospects.
    
    \begin{table}[ht]
        \centering
        \caption{Key Information on Major VLBI Software Packages. It details critical attributes of each software, including programming language, open-source status, preprocessing capabilities, and support for single solutions (Ss), global solutions (Gs), and combined solutions (Cs).}
        \begin{tabular}{lcccccc}
            \hline
             & Language   & Open source   & PreProcess   & Ss    & Gs    & Cs    \\ \hline
            ivg::ASCOT\citep{Artz2016} & C++     & \ding{51}    & \ding{51}   & \ding{51}  & \ding{51}    & \ding{51} \\
            Calc/Solve & Fortran & \ding{51}    & \ding{51}   & \ding{51}  & \ding{51}    & \ding{55} \\
            DOGS-RI\citep{Kwak2017}    & Fortran & \ding{55}    & \ding{51}   & \ding{51}  & \ding{51}    & \ding{51} \\
            nuSolve\citep{Bolotin2014}    & C++     & \ding{51}    & \ding{51}   & \ding{51}  & \ding{55}    & \ding{55} \\
            PORT\citep{Schuh2021}       & MATLAB  & \ding{51}    & \ding{51}   & \ding{51}  & \ding{51}    & \ding{51} \\
            VTD/Psolve     & Fortran & \ding{51}    & \ding{51}   & \ding{51}  & \ding{51}    & \ding{55} \\
            Quasar     & Fortran & \ding{55}    & \ding{51}   & \ding{51}  & \ding{51}    & \ding{55} \\
            VieVS\citep{johannes2018}      & MATLAB  & \ding{51}    & \ding{55}   & \ding{51}  & \ding{51}    & (Planned) \\
            Where\citep{Kirkvik2017}      & Python  & \ding{51}    & \ding{55}   & \ding{51}  & \ding{51}    & (Planned) \\
            GASV       & Python  & \ding{51}    & \ding{51}   & \ding{51}  &    \ding{51}    & (Planned) \\ \hline
        \end{tabular}
        \label{analysis_software}
    \end{table}

    \section{Structure and Algorithm} \label{sec2}
    \subsection{Structure}
\begin{figure}[ht]
        \centering
        \includegraphics[width=0.7\textwidth]{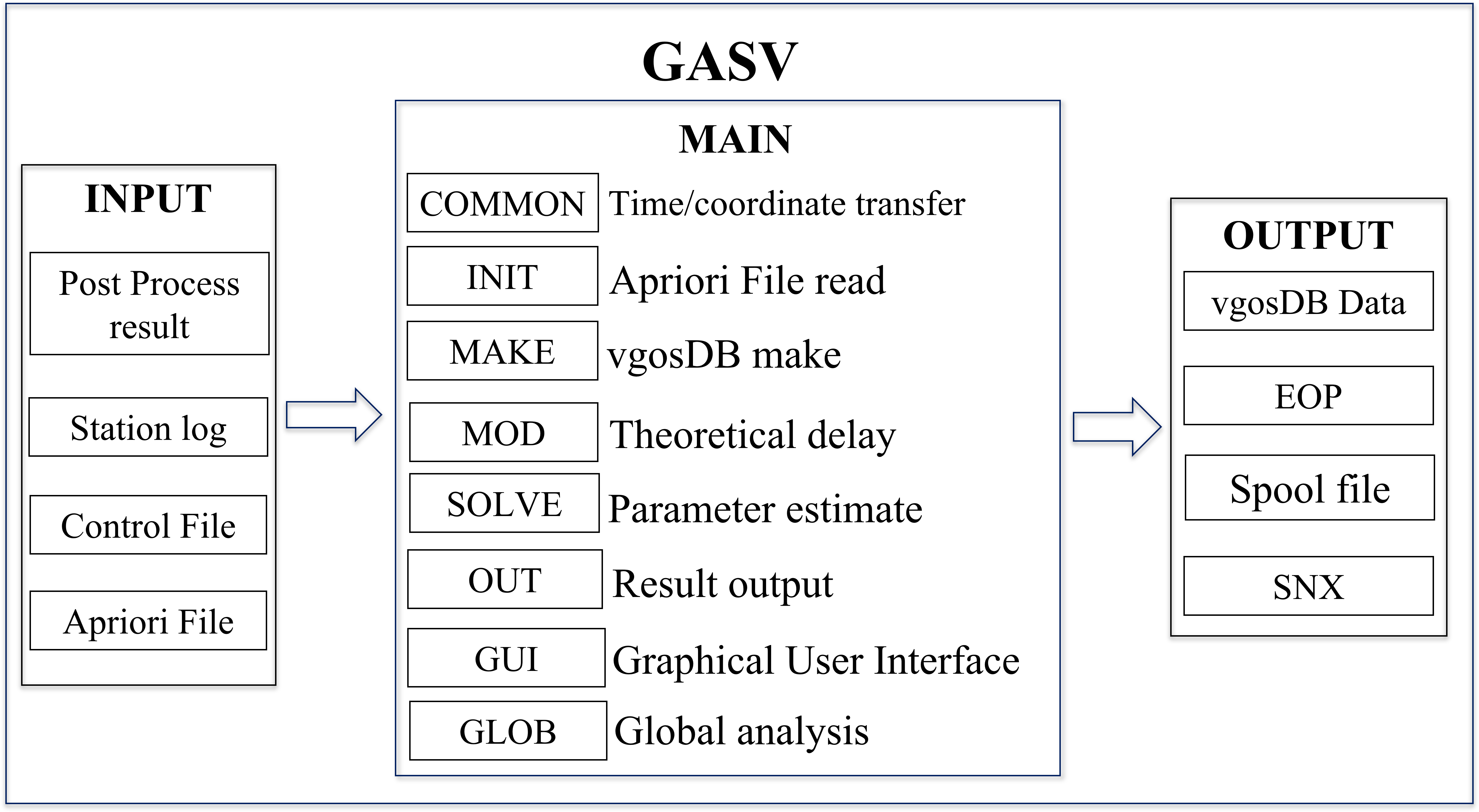}
        \caption{Structure of GASV software. The workflow begins with INPUT data, including the Post-Process results (MarkIV/NGS), station log, control file, and a priori file. These inputs are processed by the MAIN module. The OUTPUT module produces four result types: vgosDB, EOP files, Spool files, and SINEX files}
        \label{Fig_software}
    \end{figure}

    GASV is developed in Python—a high-level programming language widely adopted in data science and machine learning for its flexibility and readability. Figure \ref{Fig_software} illustrates the structure of GASV, which consists of three parts: INPUT, MAIN and OUTPUT. 
    
    All data-related input files, control files, a priori files and log files are saved in INPUT. The MAIN part contains the core code, organized into approximately 100 decoupled functions and procedures grouped into seven folders, inspired by VieVS. Time and coordinate transformation functions are stored in the COMMON directory. INIT includes functions for reading a priori files and initializing parameters. MAKE contains functions for creating the vgosDB database. Procedures for calculating theoretical delay, atmospheric delays and ionospheric delays are located in the MOD directory. Codes used for least-squares fitting and parameter estimation are saved in the SOLVE directory. The GUI includes codes for the Graphical User Interface. GLOB includes functions and procedures used for global analysis to generate the Terrestrial Reference Frame (TRF) and Celestial Reference Frame (CRF). Procedures for exporting results (vgosDB, EOP files\footnote{https://ivscc.gsfc.nasa.gov/IVS\_AC/files\_IVS-AC/}, Spool files and Solution Independent Exchange format (SINEX) files\footnote{https://www.iers.org/IERS/EN/Organization/AnalysisCoordinator/SinexFormat/sinex.html}) are saved in OUT. All output files are saved in the OUTPUT directory.

    \subsection{VLBI Delay Model}
    VLBI measures the time difference ($\tau$) of the radio signal arrival from distant quasars at two Earth-based antennas, defined as the VLBI baseline delay. The VLBI delay is modeled as follows:
    \begin{equation}
        \centering
        \tau_{observe}=\tau_{theory} + \Delta \tau_{clock} + \Delta \tau_{trop} + \Delta \tau_{iono} + \Delta \tau_{device} + \Delta \tau_{cable}
    \end{equation}
    where $\tau_{observe}$ is the observed delay, and $\tau_{theory}$ is the theoretical geometric delay, which accounts for gravitational delay caused by solar system bodies (e.g., the Earth, the Sun) and relativistic effects. Detailed calculation of $\tau_{theory}$ follows the IERS Conventions (2010) \citep{iers2010}. $\Delta \tau_{clock}$ is the correction term for clock errors between the different H-maser clocks used at two stations. $\Delta \tau_{trop}$ and $\Delta \tau_{iono}$ correct for tropospheric and ionospheric delays as signals pass through the atmosphere, $\Delta \tau_{device}$ accounts for device delays, including the thermal/gravitational deformation \citep{Wresnik2007,Nothnagel2019}, axis offsets \citep{Nilsson2017}, and channel delays. $\Delta \tau_{cable}$ accounts for cable delay due to different cable lengths \citep{Pablo2022}. In GASV, all delay components except the ionospheric delay (calibrated using multi-band group delay, see Section \ref{sub_ion}) and cable delay are fully modeled.

    \subsection{Ambiguity Resolution Algorithm}
    Station-based instrumental phases, baseline delays (frequency slope of phase), and delay rates (time slope of phase) can be fitted from the phases of visibility data using the fringe fitting procedure. Due to bandwidth synthesis, the group delay may contain an ambiguity equal to the inverse of the bandwidth of a single baseband channel \citep{Rogers1970}. When three stations simultaneously observe the same target, the closure residual of the group delay around the triangle formed by the three baselines should be zero. This property can be used to resolve the baseline ambiguities. Figure \ref{Fig_closure} schematically illustrates baseline group delay closure in a four-station network. Assuming ambiguities exist on the baselines, the group delay closure residuals for triangular sub-networks (e.g., ABC, ABD, ACD, BCD) are given by:
    
    \begin{figure}[ht]
        \centering
        \includegraphics[width=0.7\textwidth]{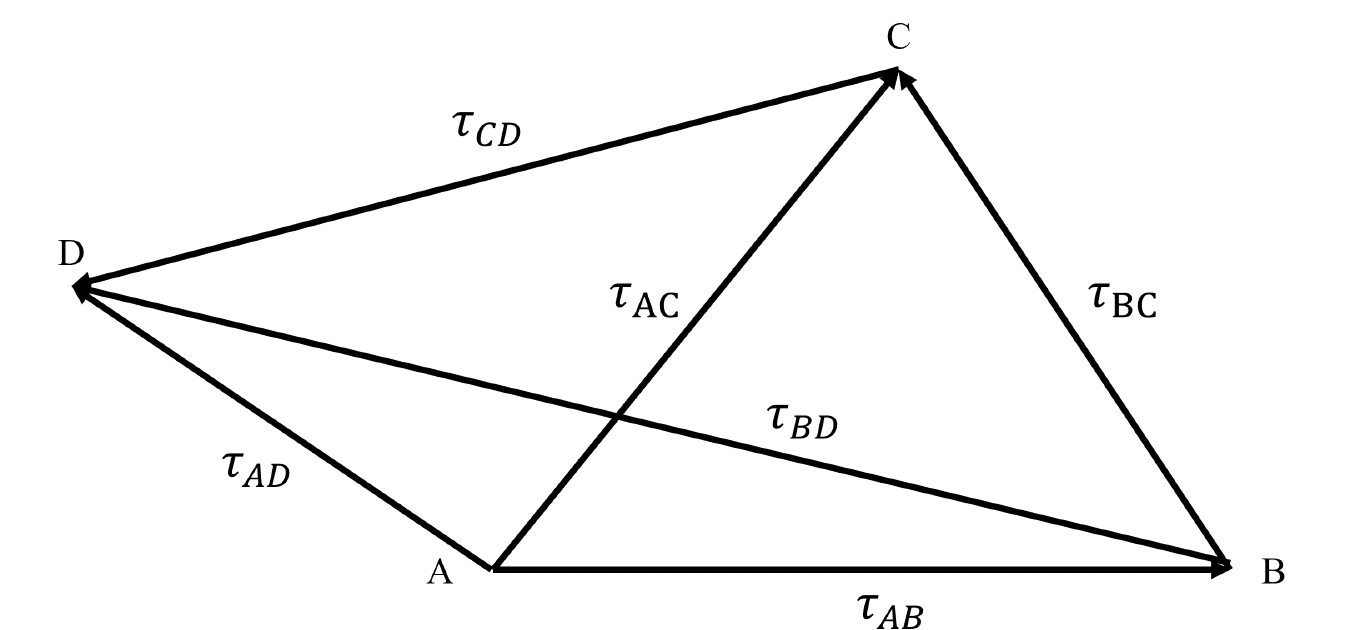}
        \caption{\label{Fig_closure} {Schematic illustration of closure delay. Letters A, B, C and D denote four stations, while $\tau$ represents the baseline delay.}}
    \end{figure}
    
    \begin{equation}
        \centering
        \tau_{AC}+N_{AC} \times A_{AC}=(\tau_{AB}+N_{AB} \times A_{AB})+(\tau_{BC}+N_{BC} \times A_{BC})
    \end{equation}
    
    \begin{equation}
        \centering
        \tau_{AD}+N_{AD} \times A_{AD}=(\tau_{AC}+N_{AC} \times A_{AC})+(\tau_{CD}+N_{CD} \times A_{CD})
    \end{equation}
    
    \begin{equation}
        \centering
        \tau_{BD}+N_{BD}\times A_{BD}=(\tau_{BC}+N_{BC} \times A_{BC})+(\tau_{CD}+N_{CD} \times A_{CD})
    \end{equation}
    
    \begin{equation}
        \centering
        \tau_{AD}+N_{AD} \times A_{AD}=(\tau_{AB}+N_{AB}\times A_{AB})+(\tau_{BD}+N_{BD} \times A_{BD}) 
    \end{equation}
    
    Here, $A_{XY}$ (ambiguity spacing) equals $1/B_{XY}$, where $B_{XY}$ is the single-channel bandwidth for the baseline between station $X$ and station $Y$, and $N_{XY}$ is the integer ambiguity number. To estimate $N_{XY}$, station A (the highest-quality station in the network, forming baselines with all others) is designated as the reference station, with N$_{A*}$ = 0 set for all baselines originating from A. The least-squares method (LSM) is then used to minimize the residual vector $v$ (Equation \ref{eq5}) and estimate $N_{XY}$ for the remaining baselines:
    
    \begin{equation}
        \centering
        \mathbf{\upsilon}  =\mathbf{A} \mathbf{x} -\mathbf{y}  \label{eq5}
    \end{equation}
    
    where, 
    \begin{equation}
        \centering
        \mathbf{y} =\begin{bmatrix}
        \tau_{AC}-\tau_{AB}-\tau_{BC}  \\
        \tau_{AD}-\tau_{AC}-\tau_{CD} \\
        \tau_{BD}-\tau_{BC}-\tau_{CD} \\
        \tau_{AD}-\tau_{AB}-\tau_{BD}
        \end{bmatrix},\mathbf{A}=\begin{bmatrix}
        A_{BC}  & 0 & 0\\
        0  & A_{BD} & 0\\
        -A_{BD}  & A_{BC} & A_{CD}\\
        0  & A_{BD} & 0
        \end{bmatrix},\mathbf{x} = \begin{bmatrix}
        N_{BC} \\
        N_{BD} \\
        N_{CD}
        \end{bmatrix}
    \end{equation}
    
    \subsection{Ionospheric Correction}\label{sub_ion}
    
    For S/X band observations, the ionospheric delay of the X band can be calculated from S and X band group delays using Equation (\ref{eq7}):
    
    \begin{equation}
        \centering
        \tau_{igx}=\frac{f_{gs}^{2}}{f_{gx}^{2}-f_{gs}^{2}}(\tau_{gx}-\tau_{gs}) \label{eq7}
    \end{equation}
    
    where $f_{gx}$ and $f_{gs}$ are the observed frequencies of X and S bands, $\tau_{gx}$ and $\tau_{gs}$ are the group delay of $X$ and $S$ bands. Ionospheric delay calibration for VLBI Global Observing System (VGOS) data differs from legacy S/X band data \citep{Nilsson2023}. For VGOS observations with 4$\times$0.5 GHz data, the differential total electron content ($\Delta TEC$) can be directly estimated during the post-processing fringe fitting based on the relationship between baseline phase, frequency and $\Delta TEC$ \citep{Huang2021}:
    
    \begin{equation}
    \centering
    \phi (f)=\tau _{g} \cdot (f-f_{0})+\phi _{0}-1.3445/f\cdot \Delta TEC \label{eq8}
    \end{equation}
    
    where $\phi (f)$ is the baseline phase, $f$ is observational frequency of frequency channels, $f_{0}$ is the reference frequency, and $\phi _{0}$ is the initial phase.
    
    \subsection{Outlier Flagging and Reweighting}

    Outliers are common in VLBI data analysis and must be identified and removed to ensure result accuracy. GASV uses the $3\sigma$ criterion for outlier detection, defined as follows:
    
    \begin{equation}
        \centering
        WRMS_{bl}=\sqrt{\frac{V_{bl}^{T}P_{bl}V_{bl}}{\sum P_{bl}} }
    \end{equation}
    
    \begin{equation}
        \centering
        \begin{cases}
            if\quad \tau _{ibl} >3\ast WRMS_{bl}, \quad remove \\
            if\quad \tau _{ibl} \le 3\ast WRMS_{bl}, \quad keep
        \end{cases}
    \end{equation}
    
    Here, $bl$ denotes the baseline, $V_{bl}$ is the baseline residual delay, $P_{bl}$ is the baseline weight, and $WRMS_{bl}$ is the weighted root mean square (WRMS) of the residual delay. GASV can automatically identify and flag the data points where the post-fit residual delay exceeds three times the WRMS. It also supports interactive manual editing.

\begin{figure}[ht]
        \centering
        \includegraphics[width=0.7\textwidth]{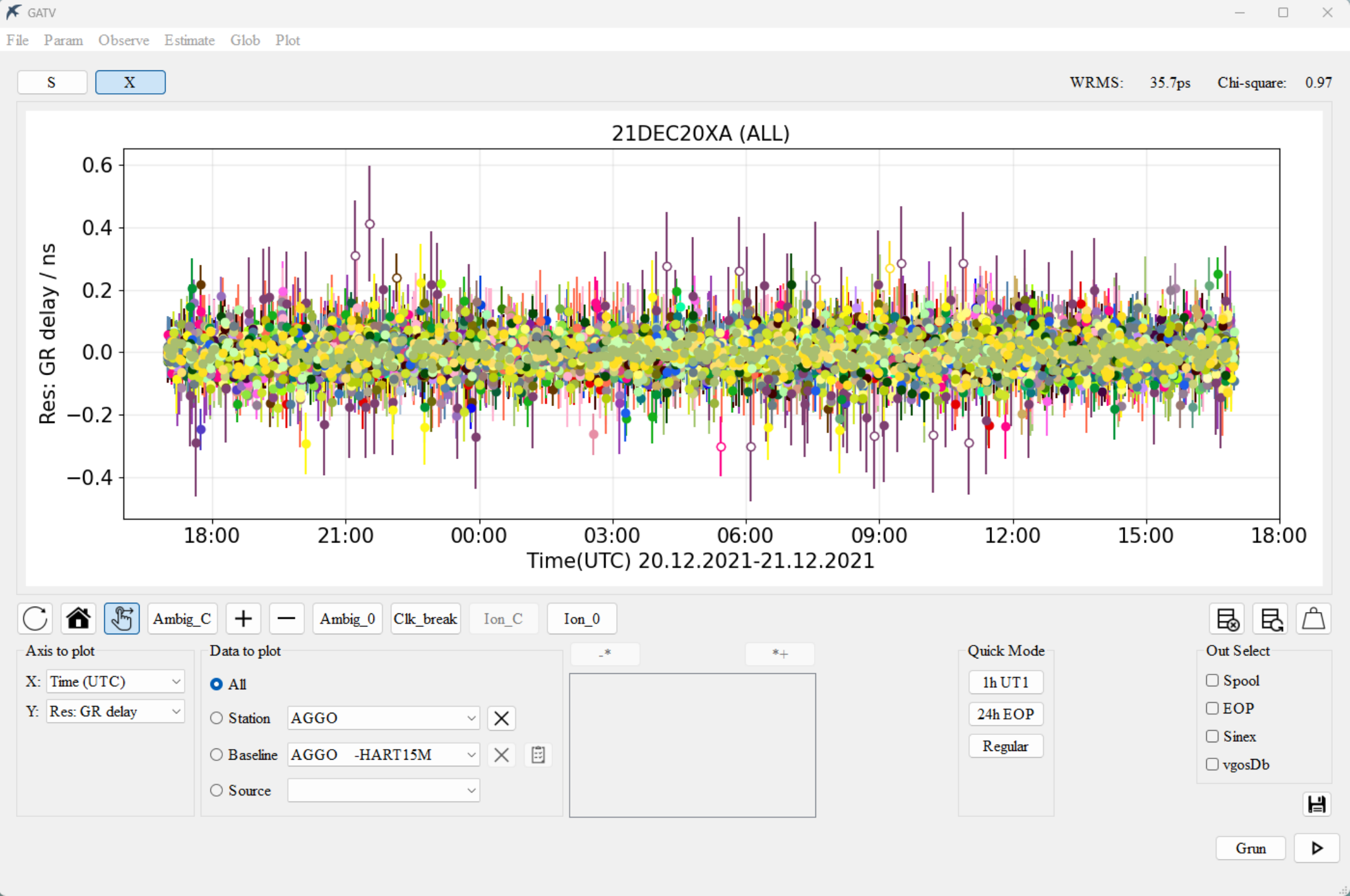}
        \caption{Software screenshot showing observation residuals of 21DEC20XA session with outlier selection. The software supports residual visualization by station, baseline, and source; it also allows for quick configuration of estimation parameters and saves processing results.}
        \label{Fig_GUI_outlier}
    \end{figure}

    Figure \ref{Fig_GUI_outlier} shows the interactive GUI for outlier flagging. Bad data points can be selected by dragging a region with the left mouse button, as illustrated by the hollow circle in the figure. Most data operations, such as ambiguity correction, ionospheric correction, and baseline clock difference estimation selection, can be performed on this interface.
    
    \subsection{Reweighting}
    Initially, the group delay uncertainties obtained from fringe fitting are used as the weights for data analysis. However, the goodness-of-fit ($\chi^{2}$) often deviates from 1, indicating a mismatch between the initial uncertainties and the estimated parameter uncertainties. This necessitates reweighting to improve result reliability. Two reweighting modes are commonly used: station-based and baseline-based. GASV adopts the baseline-based approach, with reweighting formulas given by Equations (\ref{reweight}) and (\ref{iterations}):
    
    \begin{equation}
        \centering
        q_{i}=\sqrt{\frac{\sum(R_{i})-(n_{i}-\sum(VA^{T}_{i}w^{-1}_{i}A_{i}))}{\sum(w^{-1}_{i})-\sum(VA^{T}_{i}w^{-2}_{i}A_{i})}}
        \label{reweight}
    \end{equation}
    
    \begin{equation}
        \centering
        w_{new} = w_{old}+\sum\limits_{i}q^{2}_{i}\textbf{I}
        \label{iterations}
    \end{equation}
    
    Here, $i$ represents the baseline index, $R$ is the vector of post-fit residuals, $n$ is the number of observations, $V$ is the covariance matrix of the estimates, $A$ is the matrix of condition equations, and $w$ is the weight matrix.
    
    \subsection{Session Analysis}
    For single-session data analysis, a priori station motions are calculated based on the positions and velocities of stations within the ITRF frame defined by the a priori file. Site displacement caused by solid Earth tides and ocean tides are calculated following the IERS Conventions (2010). The theoretical delay, including relativistic corrections, is implemented according to the IERS Conventions (2010). Earth rotation variations due to zonal tidal variations with periods from 5 to 18.6 years are modeled as specified in the IERS Conventions (2010). EOPs can be used in formats compatible with the IERS C04 series\footnote{https://www.iers.org/IERS/EN/DataProducts/EarthOrientationData/eop.html} or the US Naval Observatory (USNO) finals series\footnote{https://cddis.nasa.gov/archive/vlbi/gsfc/ancillary/solve\_apriori}. Partial derivatives of the VLBI observation equation with respect to key geodetic parameters are referenced from \cite{Kamil2012}. For the EOP model, additional subdaily variations are incorporated following the IERS Conventions (2010), based on the method proposed by \cite{Desai2016}.
    
    Parameter adjustment is currently based on the classical weighted least-squares method. For sub-daily parameter estimation, continuous piecewise linear offset (CPWLO) functions are employed. Polar motion and UT1 can be estimated as CPWLO functions, or by calculating their offsets and rates at specific epochs (e.g., midnight, noon, or the session midpoint). In addition to EOP, station positions, radio source coordinates, and temporal parameters-including zenith tropospheric wet delay, station clock offsets and baseline clock offsets-are also determined.
    
    \subsection{Global Analysis}
    
    Global analysis of long-term historical data is essential for maintaining VLBI station position (coordinate and velocity), monitoring the crustal motion of the solid Earth, and establishing high-precision CRF and TRF. The GLOB module as shown in Figure \ref{Fig_software} is used for global analysis. 
    
    In global analysis, single-epoch normal equations are first reconstructed from SINEX files. Local parameters (e.g., EOP) are reduced, while global parameters-specifically radio source positions and station positions-are retained. Station positions are then linearly modeled using coordinates at a reference epoch and long-term station velocities. Subsequently, the normal equations from each epoch are stacked. Finally, No-Net-Rotation (NNR) and No-Net-Translation (NNT) constraints are applied to station positions, velocities, and radio source positions. In addition, constraints on the relative velocities of co-located stations are applied. These constraints are used to form the final multi-epoch normal equation system for parameter estimation. Detailed matrix operations for global analysis are provided in Appendix \ref{app1}.
    
    \section{Performance evaluation} \label{sec3}
    To validate GASV’s performance and accuracy, we processed data from the International VLBI Service for Geodesy and Astrometry (IVS) and compared the results with those from two leading IVS analysis centers: BKG (Federal Agency for Cartography and Geodesy) and USNO.
    \subsection{Parameter Setting}
    
    Station coordinates were fixed to ITRF2020. Source coordinates were fixed to the 3rd realization of the ICRF (ICRF3). Regarding station coordinate correction, we adopt the IERS Conventions (2010), that including solid earth tides, tidal ocean loading (TPXO72), pole tide, ocean pole tide and atmospheric tide loading, while non-tidal atmospheric loading was not taken into account. For the priori EOP, the USNO EOP file was used and the high frequency (diurnal and subdiurnal) corrections were compensated in polar motion and UT1. The Global Mapping Function (GMF) for tropospheric delay was used. Table \ref{apriori_set} lists the priori models used in GASV, Table \ref{solve_set} is the detail setup for one-hour observation and 24-hour observation used in this section for performance evaluation of GASV. For the 24-hour solutions, three modes are used, Mode 1 estimates EOP only, Mode 2 estimates EOP, station positions and radio source positions, Mode 3 is designed for sub-daily EOP estimation.
    
    \begin{table}[htbp]
        \centering
        \caption{Apriori models used in session analysis}
        \begin{tabular}{lll}
        \hline
        Station priori position      & \multicolumn{2}{l}{ITRF2020}              \\
        Radio source priori position & ICRF3                          &          \\
        Solid Earth tides             & IERS Conventions (2010)          &          \\
        Ocean pole tide loading       & \multicolumn{2}{l}{IERS Conventions (2010)} \\
        Ocean tidal loading           & \multicolumn{2}{l}{TPX072\citep{Egbert2002}} \\
        Atmospheric tidal             & \multicolumn{2}{l}{IERS Conventions (2010)} \\
        Daily EOPs                    & \multicolumn{2}{l}{'usno\_finals.erp'}      \\
        Subdaily EOP model            & \multicolumn{2}{l}{Desai model}           \\
        Precession/nutation model     & \multicolumn{2}{l}{IAU 2006/2000A}        \\
        Tropospheric mapping function  & \multicolumn{2}{l}{GMF}                   \\ \hline
        \end{tabular}
        \label{apriori_set}
    \end{table}
    
    \begin{table}[htbp]
        \centering
        \caption{Parameter Estimation Configurations for Different Session Types. For 1-hour session, only one configuration exists. For 24-hour session, three modes exist: all modes share the same settings for station clock, zenith wet delay, and gradient; Mode 1 and Mode 2 share identical configurations for UT1, LOD, polar motion, polar motion rate, and nutation. Mode 2 and Mode 3 share the same settings for station position, source position, and datum constraints.}
        \begin{tabular}{l|c|ccc}
        \hline
        \multirow{2}{*}{\textbf{Estimate Parameter}} & \multirow{2}{*}{\textbf{1-hour session}}     & \multicolumn{3}{c}{\textbf{24-hour session}}                                                                                                                      \\ \cline{3-5} 
                                                     &                                                                          & Mode 1                                         & Mode 2                                       & Mode 3                                                                      \\ \hline
        Clocks                                       & \begin{tabular}[c]{@{}c@{}}Interval:1h\\ Constraint: 1.3 cm\end{tabular} & \multicolumn{3}{c}{\begin{tabular}[c]{@{}c@{}}Interval: 1 h\\ Constraint: 1.3 cm\end{tabular}}                                                                            \\
        Zenith wet delay                             & \begin{tabular}[c]{@{}c@{}}Interval:1h\\ Constraint:1.5 cm\end{tabular}  & \multicolumn{3}{c}{\begin{tabular}[c]{@{}c@{}}Interval: 1 h\\ Constraint: 1.5 cm\end{tabular}}                                                                            \\
        troposphere gradients                        & NO                                                                       & \multicolumn{3}{c}{\begin{tabular}[c]{@{}c@{}}Interval: 6 h\\ Constraint: 0.5 mm\end{tabular}}                                                                            \\
        UT1                                          & \begin{tabular}[c]{@{}c@{}}offset\\ Constraint:3 ms\end{tabular}         & \multicolumn{2}{c}{\begin{tabular}[c]{@{}c@{}}offset\\ Constraint: 3 ms\end{tabular}}       & \begin{tabular}[c]{@{}c@{}}Interval: 1 h\\ Constraint: 3 ms\end{tabular}    \\
        LOD                                          & NO                                                                       & \multicolumn{2}{c}{\begin{tabular}[c]{@{}c@{}}offset\\ Constraint: 3 ms/day\end{tabular}}   & NO                                                                         \\
        Polar motion                                 & NO                                                                       & \multicolumn{2}{c}{\begin{tabular}[c]{@{}c@{}}offset\\ Constraint: 10 mas\end{tabular}}     & \begin{tabular}[c]{@{}c@{}}Interval: 1 h\\ Constraint: 10 mas\end{tabular}  \\
        Rate of Polar motion                         & NO                                                                       & \multicolumn{2}{c}{\begin{tabular}[c]{@{}c@{}}offset\\ Constraint: 10 mas/day\end{tabular}} & NO                                                                         \\
        Nutation                                     & NO                                                                       & \multicolumn{2}{c}{offset}                                                                  & offset                                                                     \\
        Station position                             & NO                                                                       & NO                                            & \multicolumn{2}{c}{offset}                                                                                               \\
        Radio source position                        & NO                                                                       & NO                                            & \multicolumn{2}{c}{offset}                                                                                               \\
        CRF Datum                                    & NO                                                                       & NO                                            & \multicolumn{2}{c}{NNR to ICRF3 defining sources}                                                                        \\
        TRF Datum                                    & NO                                                                       & NO                                            & \multicolumn{2}{c}{NNR and NNT on coordinates}                                                                               \\ \hline
        \end{tabular}
        \label{solve_set}
    \end{table}
    
    \subsection{IVS Intensive Sessions}
    In order to rapidly measure the UT1, the IVS organizes one-hour intensive sessions using the east-west baselines. Currently, there exists three intensive networks, sessions from Wednesday to Sunday (INT1), sessions from Monday to Sunday (INT2) and sessions on Tuesday (INT3). We analyzed IVS intensive session data collected from 2020 to 2023, including 741 INT1 sessions and 321 INT2/INT3 sessions. Results from BKG and USNO were obtained from the IVS data center\footnote{\it https://cddis.nasa.gov/archive/vlbi/ivsproducts}, while NTSC results were derived using GASV.
    
\begin{figure}[ht]
        \centering
        \includegraphics[width=0.4\textwidth]{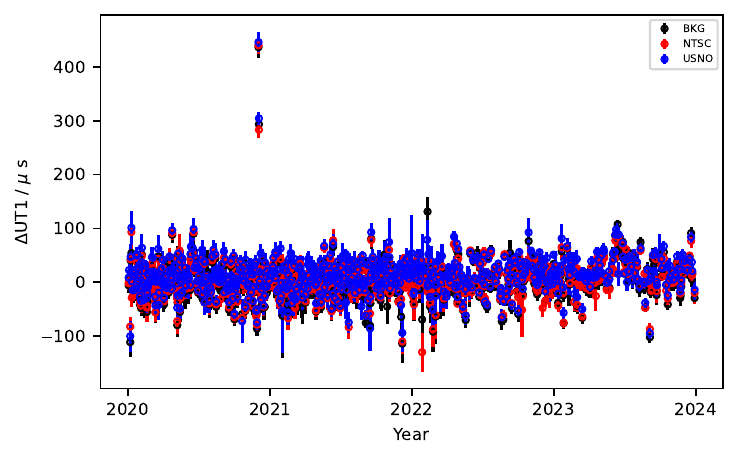}
        \includegraphics[width=0.4\textwidth]{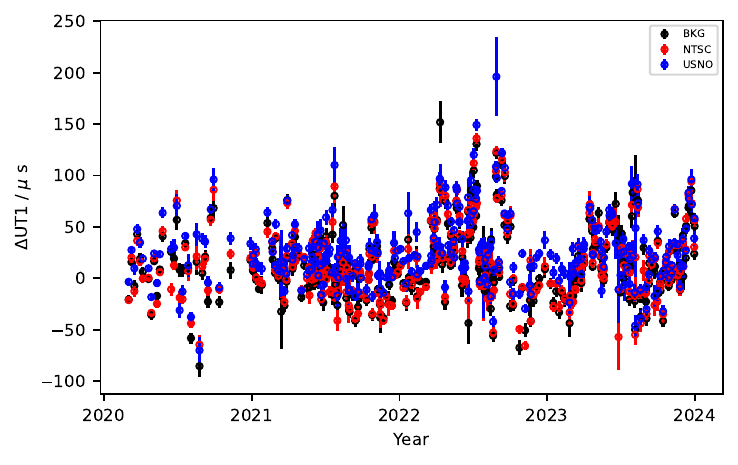}
        \caption{Comparison of the results of UT1 among different analysis centers from 2020 to 2023, w.r.t. C04. The left panel shows the INT1 result, the right panel shows the INT2 and INT3 result.}
        \label{Fig_compare_1h}
    \end{figure}

    Figure \ref{Fig_compare_1h} displays the differences in UT1-UTC ($\delta$UT1) for solutions from BKG, USNO, and NTSC, relative to the IERS C04 20 series\footnote{\it https://datacenter.iers.org/data/latestVersion/EOP\_20\_C04\_one\_file\_1962-now.txt}. Here, $\Delta UT1=\delta UT1_{AC}-\delta UT1_{C04}$, where AC denotes the analysis center abbreviation, and $\delta UT1_{C04}$ is the UT1 value interpolated from C04 series. The left panel shows the results for INT1 sessions, while the right panel shows the results for INT2/INT3 sessions. It can be seen that the NTSC results calculated with GASV are highly consistent with those from BKG and USNO.
    
    Table \ref{table_int1_result} summarizes the statistics of $\Delta UT1$ (including mean offset, RMS of offset, and mean formal error) for different analysis centers. NTSC’s INT session results are on par with BKG’s, with slightly lower mean and RMS values than USNO’s. This difference is primarily attributed to the use of different a priori station positions: NTSC and BKG both used ITRF2020, while USNO used its own TRF.
    
    \begin{table}[htbp]
        \centering
        \caption{Statistics of $\Delta UT1$ for different analysis center ($\mu s$)}
        \begin{tabular}{lccccc}
        \hline
        Type                        & \multicolumn{1}{l}{Session Number}         & AC   &  Mean Offset & RMS of Offset & Mean Formal error \\ \hline
        \multirow{3}{*}{INT1}       & \multirow{3}{*}{741}   & NTSC &  3.6         & 34.5          & 12.4              \\
                                    &                                         & BKG  &  1.7         & 35.4          & 14.2              \\
                                    &                                         & USNO &  12.7        & 37.1          & 14.4              \\ \hline
        \multirow{3}{*}{\begin{tabular}[c]{@{}l@{}}INT2\\ INT3\end{tabular}} & \multirow{3}{*}{321}   & NTSC &  15.6        & 37.8          & 7.3               \\
                                                                             &                        & BKG  &  15.0        & 37.8          & 7.6               \\
                                                                             &                        & USNO &  26.8        & 42.5          & 7.5               \\ \hline 
        \end{tabular}
        \label{table_int1_result}
    \end{table}
    
    \subsection{IVS Regular Sessions}
\begin{figure}[ht]
        \centering
        \includegraphics[width=0.7\textwidth]{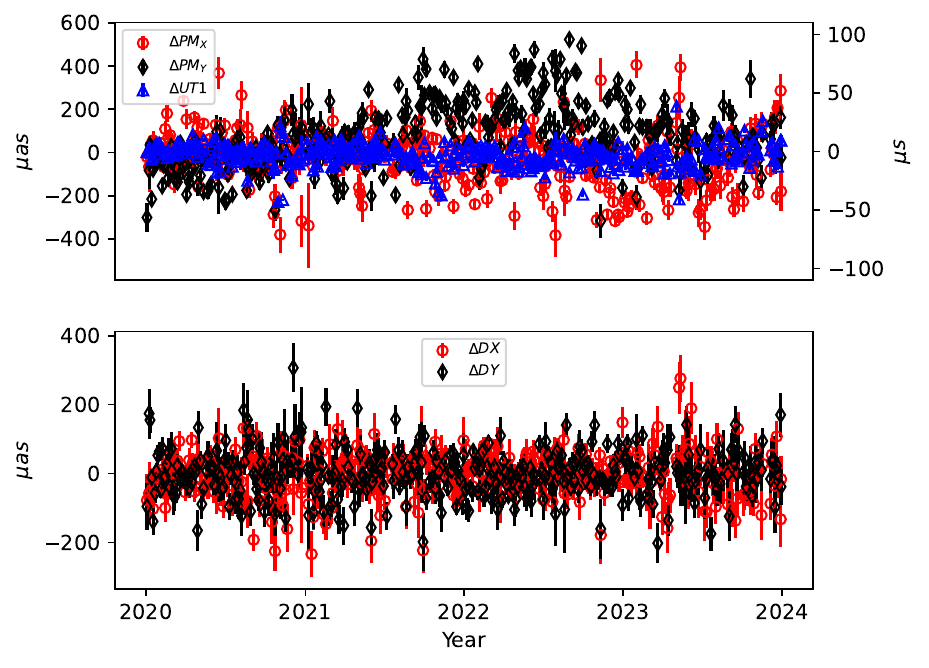}
        \caption{\label{Fig_result_compare}Estimated results of EOP from 2020 to 2024, w.r.t. C04. The upper panel presents results for polar motion and UT1, where the left Y-axis corresponds to polar motion results and the right Y-axis represents the UT1 results. The lower panel displays results for the CPO.}
        \label{fig_EOP}
    \end{figure}

   Every Monday and Thursday, the IVS organizes 24-hour observations by using 6-15 VLBI stations which are named as IVS regular sessions. Here we used IVS Regular session data from 2020 to 2024 to evaluate GASV’s performance in determining EOP and station coordinates, following the parameter settings in Table \ref{solve_set}. Figure \ref{fig_EOP} shows the estimated EOP relative to C04 from 2020 to 2024. The upper panel displays polar motion and UT1 results. The left Y-axis corresponds to polar motion components ($PM_X$ and $PM_Y$), and the right Y-axis corresponds to UT1, the $PM_{X}$ is polar motion X axis, the $PM_{Y}$ is polar motion Y axis. Variations in $\Delta PM_{X}$ are within $\pm~200$ $\mu$as. Here, we identify an obvious bias in $\Delta PM_{Y}$ during 2022-2023. We compared our results with those from other analysis centers and found that the results provided by Italian Space Agency(ASI) and Institute of Applied Astronomy (IAA) also exhibit this phenomenon. Variations in $\Delta UT1$ are concentrated within $\pm 20$ $\mu$s. Statistically, the RMS values of $\Delta PM_{X}$, $\Delta PM_{Y}$ and $\Delta UT1$ are 129.2 $\mu$as, 168.6 $\mu$as, and 10.2 $\mu$s, respectively. Their mean formal errors are 41.7 $\mu$as, 41.0 $\mu$as and 2.6 $\mu$s, respectively. The lower panel shows the Celestial Pole Offset (CPO) results. Variations in $\Delta DX$ and $\Delta DY$ are concentrated within $\pm$100 $\mu as$, with RMS values of 65.1 $\mu$as and 68.3 $\mu$as, and mean formal errors of 43.8 $\mu$as and 43.9 $\mu$as, respectively. All these statistics are within the reasonable ranges.
    
\begin{figure}[ht]
        \centering
        \includegraphics[width=0.7\textwidth]{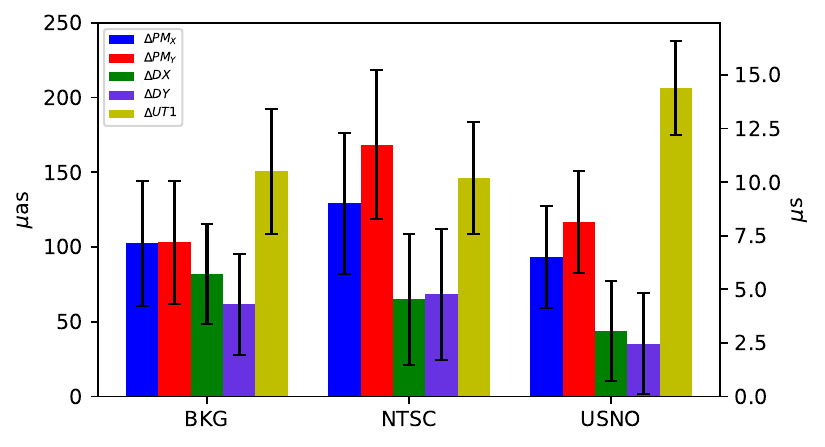}
        \caption{\label{Fig_result_compare}Comparison of RMS Values for estimated EOP from different analysis centers w.r.t. C04. The left Y-axis corresponds to polor motion and CPO results, while the right Y-axis represents the UT1 results.}
        \label{fig_result}
    \end{figure}

    Figure \ref{fig_result} shows histograms of RMS values for EOP estimates relative to C04 from different VLBI analysis centers. The black bars denote the mean formal errors of EOPs. The RMS of $\Delta UT1$ for NTSC is slightly lower than those for BKG and USNO, wherease the $\Delta PM_{X}$ and $\Delta PM_{Y}$ RMS values for NTSC are slightly higher than those for BKG and USNO. In general, the results are consistent within 1$\sigma$, confirming that NTSC’s results (derived with GASV) are comparable to those from BKG and USNO. 
    
    \subsection{IVS CONT17 Session}
    In addition to Intensive and Regular sessions, the IVS organizes continuous VLBI campaigns (CONT) approximately every three years. These campaigns typically last 15 days with 24-hour observations, and one key goal is to analyze sub-daily EOP. The most recent CONT campaign (CONT17) was conducted from November 28 to December 12, 2017\citep{Behrend2018}. Unlike previous CONT campaigns, which had only one network, CONT17 included three independent networks: two legacy networks operating at S/X band and one VGOS network in 4$\times$1GHz mode.
    
\begin{figure}[ht]
        \centering
        \includegraphics[width=0.7\textwidth]{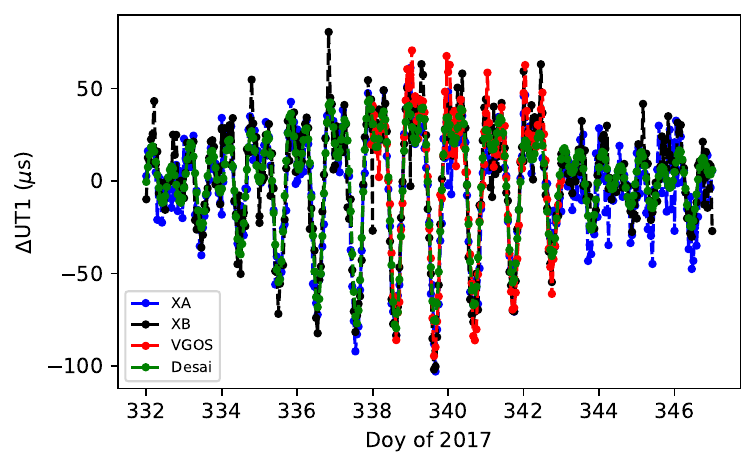}
        \caption{Comparison of hourly UT1 estimates during CONT17 w.r.t. C04. The blue dotted line represents results from the VLBA network, the black dotted line shows traditional legacy VLBI networks results, the red dotted line displays VGOS network results, the green dotted line corresponds to the Desai model.}
        \label{fig_cont17_ut1}
    \end{figure}

    Figure \ref{fig_cont17_ut1} shows the sub-daily UT1 results derived with GASV using CONT17 data. For this analysis, we used Mode 3 (Table \ref{solve_set}) and excluded the Desai model from the a priori models. The blue dotted line represents results from the VLBA network observing at S/X band (XA). The black dotted line shows results from traditional legacy VLBI networks (XB). The red dotted line displays result from the VGOS network. The green dotted line corresponds to the Desai model. The overall trends of the four lines are highly consistent, indicating that the GPS-based Desai model is highly precise and aligns well with the VLBI-derived results.
    
\begin{figure}[ht]
        \centering
        \includegraphics[width=0.7\textwidth]{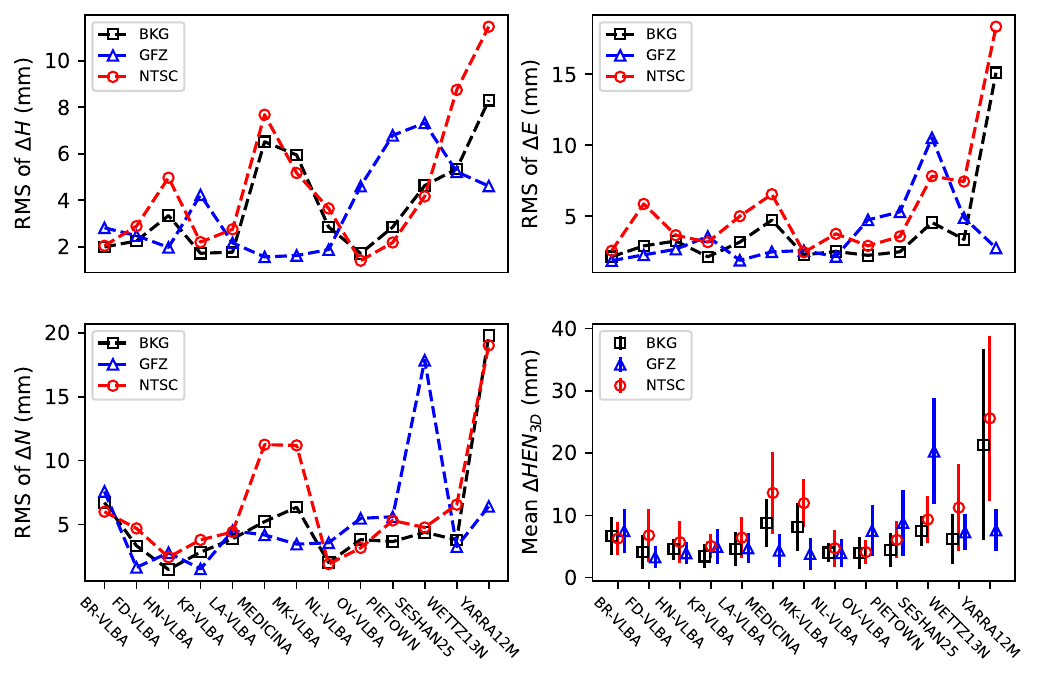}
        \caption{Station position estimates for CONT17(XA) w.r.t. ITRF2020. The panels are arranged from top-left to bottom-right as follows: horizontal, east, north and 3D position components.}
        \label{fig_cont17_station}
    \end{figure}

    Figure \ref{fig_cont17_station} shows station coordinate differences relative to ITRF2020. Results for BKG and the German Research Center for Geosciences (GFZ) were obtained from SINEX files uploaded to the IVS data center. The NTSC results were derived using GASV with the setup of Mode 2 in Table \ref{solve_set}. The upper-left panel shows the RMS of horizontal coordinates differences. The upper-right panel shows differences in the east component. The lower-left panel shows differences in the north component. 
   Equation (\ref{RMS_HEN}) is used to calculate the differences RMS in the horizontal direction over the 15-day period. The lower-right panel shows the 3D position differences, calculated using Equation (\ref{ALL_HEN}). As shown, NTSC’s results are consistent with those from BKG and GFZ, with differences at the millimeter level. 
    
    \begin{equation}
        \centering
        RMS(\Delta H_{j}) = \sqrt{\frac{\sum\limits_{i=1}\limits^{N}(H_{ij}-H_{ij_{itrf}})^2}{N}},j \: is \: station
        \label{RMS_HEN}
        \end{equation}
        
        \begin{equation}
        \centering
        \Delta HEN_{3D_{j}} = \sqrt{(H_{ij}-H_{ij_{itrf}})^2+(E_{ij}-E_{ij_{itrf}})^2+(N_{ij}-N_{ij_{itrf}})^2}
        \label{ALL_HEN}
    \end{equation}
    
\begin{figure}[ht]
        \centering
        \includegraphics[width=0.4\textwidth]{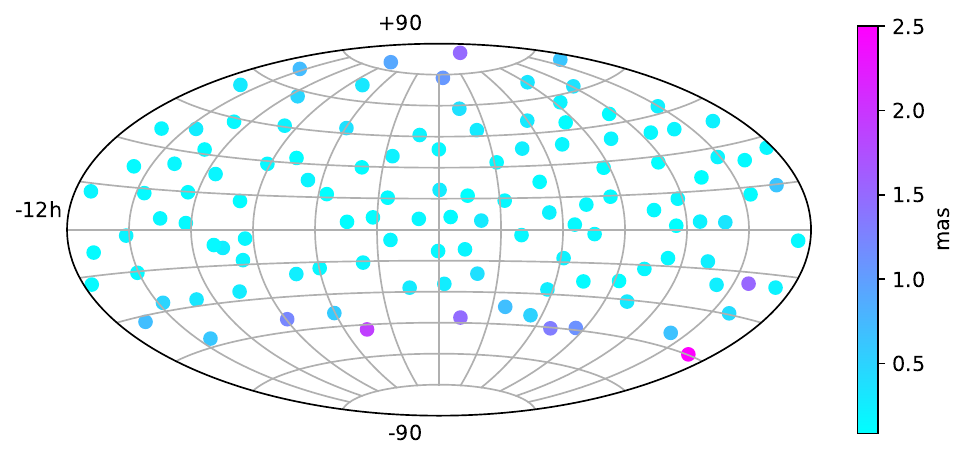}
        \includegraphics[width=0.4\textwidth]{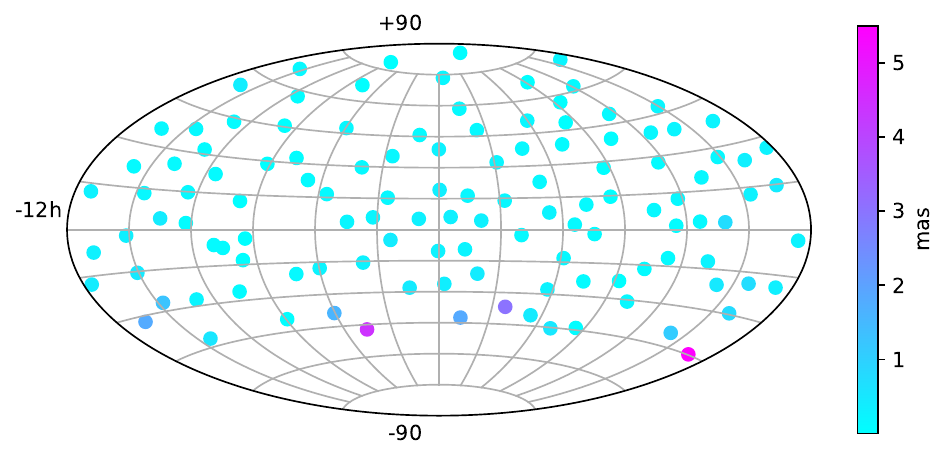}
        \caption{Differences in coordinates of right ascension (left panel) and declination (right panel)  w.r.t. ICRF3 catalog, determined by GASV using CONT17-XA data.}
        \label{fig_compare_source}
    \end{figure}

    To evaluate GASV’s astrometric capability, Figure \ref{fig_compare_source} presents the estimated right ascension and declination of 121 radio sources from CONT17-XA w.r.t. ICRF3. The left panel shows right ascension differences, and the right panel shows declination differences. Except for some sources in the southernmost region, all other sources exhibit excellent consistency with ICRF3. The RMS of the right ascension differences for the 121 sources is 0.3322 mas, and the RMS of the declination differences is 0.3597 mas.
    
    \subsection{Global Solution}
    
    We also tested GASV’s global analysis capability using its GLOB module. For this, we used SINEX files spanning 1996 to 2022, which were reduced by BKG within the ITRF2020 terrestrial reference frame and ICRF3 celestial reference frame. In total, 4,612 SINEX files were integrated, with the following constraints applied: NNR constraints on 302 defining radio sources (source 0809-493 has few observations), both NNR and NNT constraints on 55 VLBI stations, velocity field constraints on the relative motions of 13 co-located stations. Segmented processing was applied to stations with position discontinuities.
    
    \begin{figure}[ht]
        \centering
        \includegraphics[width=0.7\textwidth]{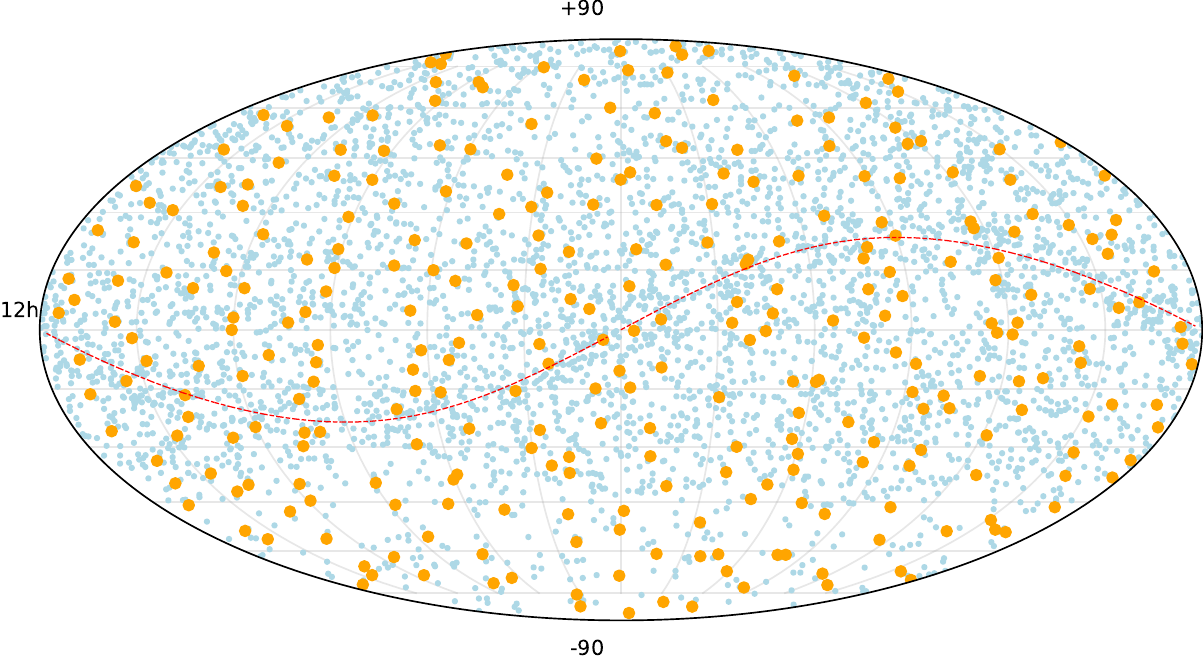}
        \caption{Locations of 4,733 radio sources from the global solution. Orange points represent defining sources, while light blue points denote non-defining sources.}
        \label{fig_compare_source_icrf}
    \end{figure}
    
\begin{figure}[ht]
        \centering
        \includegraphics[width=0.4\textwidth]{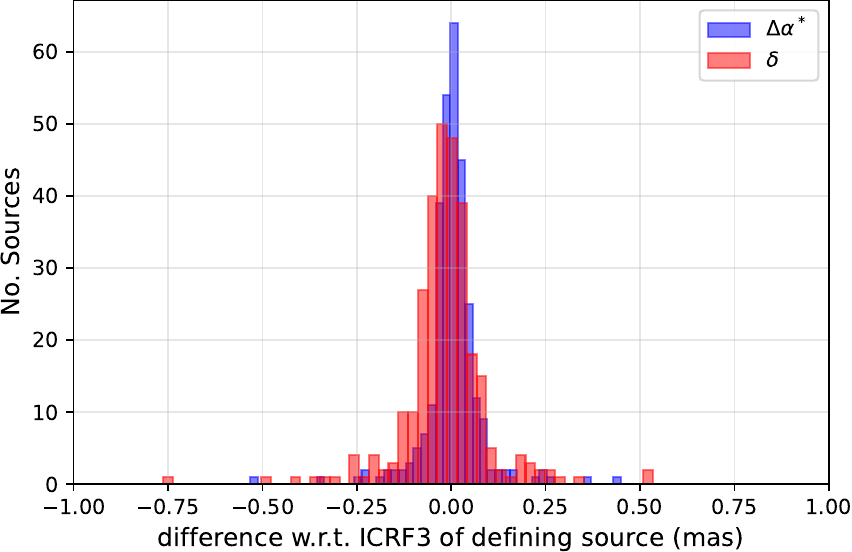}
        \includegraphics[width=0.4\textwidth]{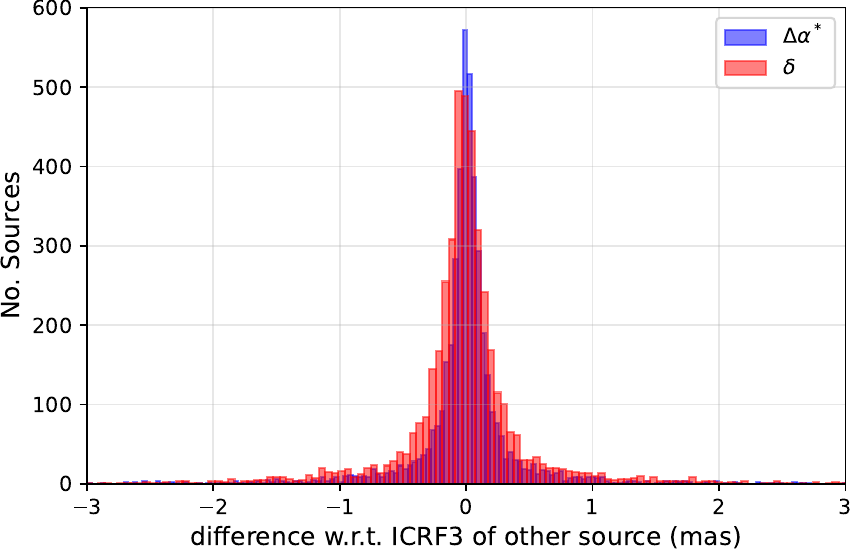}
        \caption{Histograms of coordinate differences in right ascension and declination w.r.t. ICRF3 catalog, derived from the global solution of GASV using BKG SINEX files spanning 1996 to 2022. The left panel displays results for defining sources, while the right panel shows results for non-defining sources.}
        \label{fig_compare_glob_source}
    \end{figure}

     Figure \ref{fig_compare_source_icrf} shows the locations of 4733 radio sources from global solution. Orange points represent the 302 defining sources, light blue points represent the 4431 non-defining sources, and the red line indicates the ecliptic. Figure \ref{fig_compare_glob_source} presents the statistical distributions of differences in right ascension ($\Delta\alpha ^{*}=\Delta \alpha  \cdot \cos \delta $) and declination ($\Delta\delta$) relative to ICRF3. The left panel shows the results for the defining sources, and the right panel shows results for non-defining sources. Both $\Delta\alpha ^{*}$ and $\Delta\delta$ are predominantly concentrated near zero. For defining sources, two sources (0316-444 and 0742-562) are not displayed, as their right ascension or declination differences exceed 1 mas. Sources with declination differences exceeding 0.5 mas are predominantly located in the southern hemisphere.
    
    Table \ref{table_glob_source} presents the statistical results of coordinate differences. For defining sources, the average coordinate differences are -2.8 $\mu$as in right ascension and 1.5 $\mu$as in declination, with RMS values of 0.129 mas, 0.324 mas, respectively. For non-defining sources, the averages are -13.3 $\mu$as in right ascension and -31.9 $\mu$as in declination, with RMS of 0.622 mas, 0.812 mas, respectively.
    
    \begin{table}[ht]
        \centering
        \caption{The statistics of coordinate differences as shown in Figure \ref{fig_compare_glob_source}}
        \begin{tabular}{cccccc}
        \hline
        source type                & number                   & Avg. $\Delta\alpha ^{*}$ & RMS of $\Delta\alpha ^{*}$ & Avg. $\Delta\delta$       & RMS of $\Delta\delta$    \\
                    & & ($\mu$as) & (mas) & ($\mu$as) & (mas) \\ \hline
        \multicolumn{1}{c}{defining} & \multicolumn{1}{c}{302}  & \multicolumn{1}{c}{-2.8} & \multicolumn{1}{c}{0.129}    & \multicolumn{1}{c}{1.5}   & \multicolumn{1}{c}{0.324} \\
        \multicolumn{1}{c}{non-defining}  & \multicolumn{1}{c}{4431} & \multicolumn{1}{c}{-13.3} & \multicolumn{1}{c}{0.622}    & \multicolumn{1}{c}{-31.9} & \multicolumn{1}{c}{0.812} \\ \hline
        \end{tabular}
        \label{table_glob_source}
    \end{table}
    
\begin{figure}[ht]
        \centering
        \begin{subfigure}[b]{0.6\textwidth}
            \includegraphics[width=\textwidth]{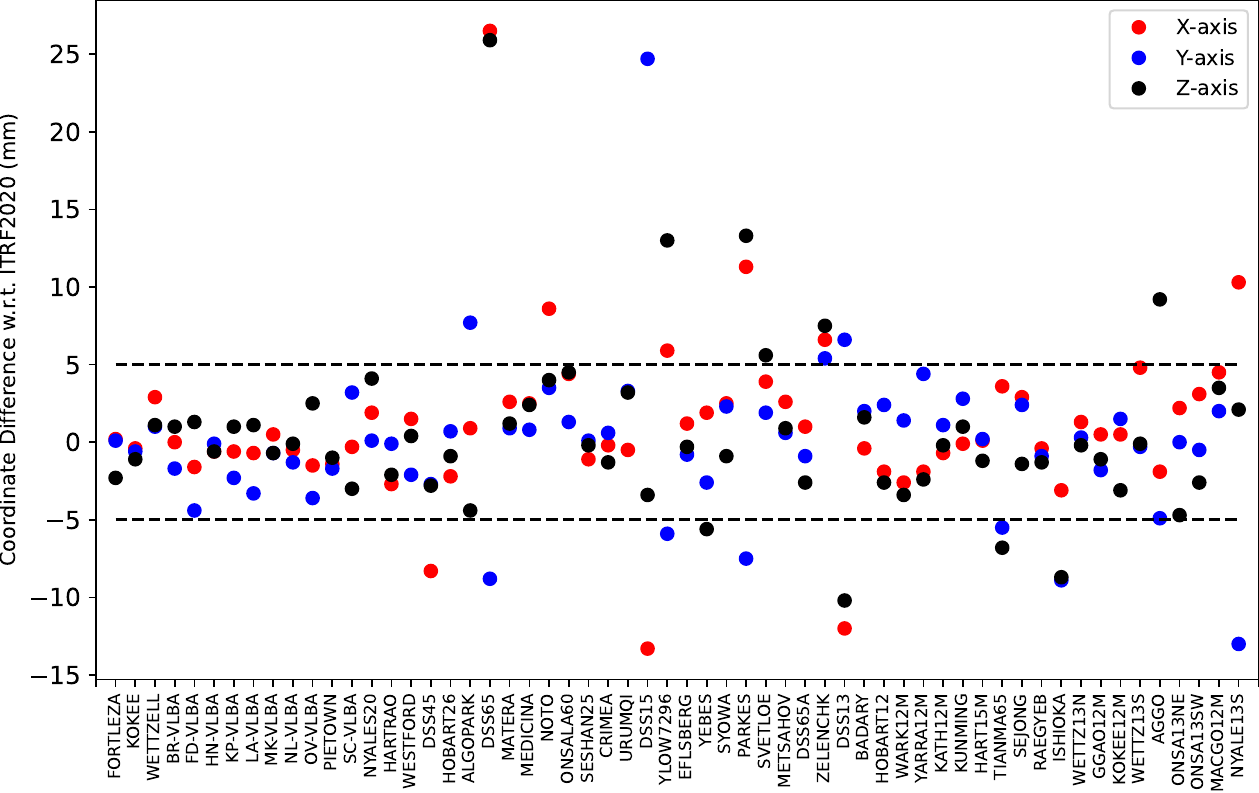}
            \caption{The differences of station coordinates} \label{fig_glob_trf:sub1}
        \end{subfigure}
        \begin{subfigure}[b]{0.6\textwidth}
            \includegraphics[width=\textwidth]{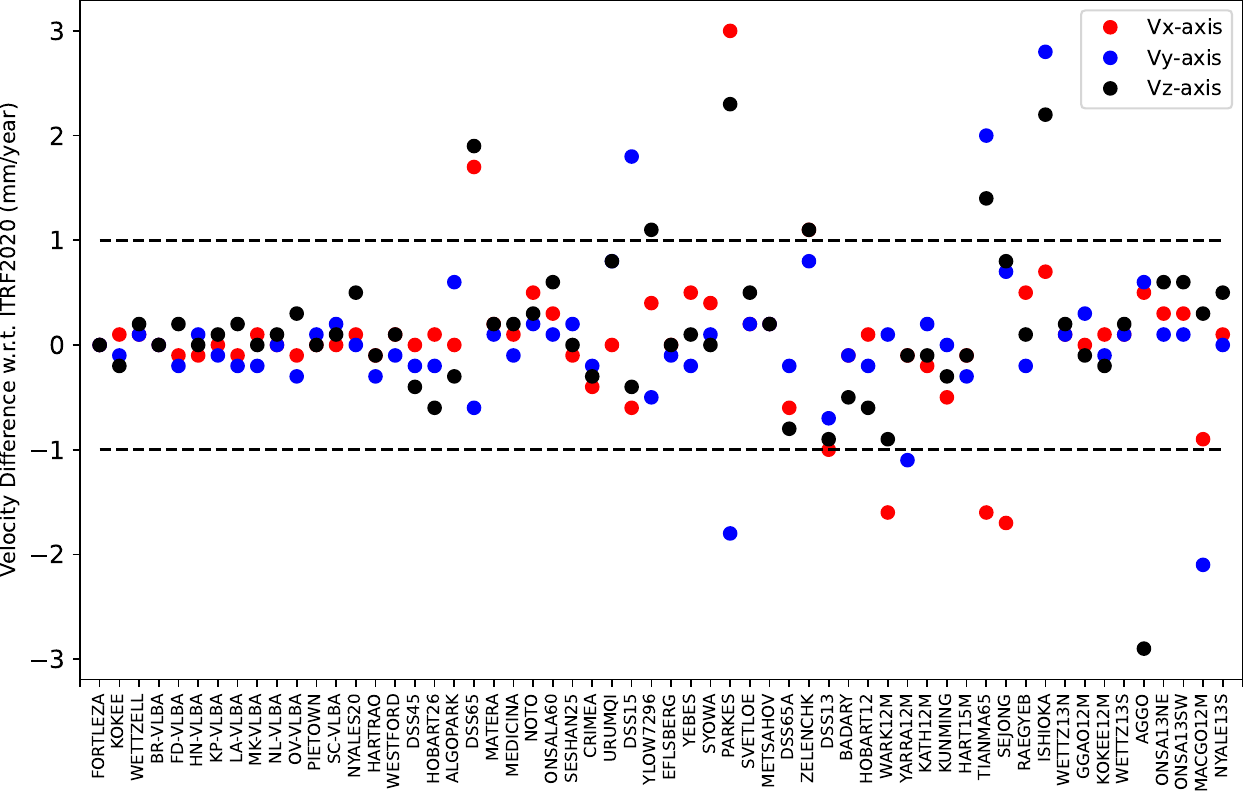}
            \caption{The differences of station velocities} \label{fig_glob_trf:sub2}
        \end{subfigure}
        \caption{Differences in station coordinates (upper panel) and station velocities (lower panel) relative to ITRF2020, derived from the global solution of GASV using BKG SINEX files spanning 1996 to 2022.}
        \label{fig_glob_trf}
    \end{figure}

    Figure \ref{fig_glob_trf} presents global analysis results for the positions and velocities of 58 VLBI stations, relative to ITRF2020. The upper panel shows station coordinate differences, and the lower panel station velocity differences. Four Stations-DSS13, DSS15, DSS65 and PARKES-participated in approximately 40 VLBI sessions over a 10-year period. Since these stations were not subject to NNR or NNT constraints, their coordinate solutions exhibit larger differences compared to constrained stations. 
    
    \section{Conclusion} \label{sec4}
    GASV is a new VLBI analysis software developed by the VLBI group at the National Time Service Center. It accepts HOPS outputs and NGS card files as input data to estimate key parameters, including EOP, sub-daily EOP, station positions, and radio source coordinates.

    To assess its performance, we conducted comprehensive tests using data from IVS intensive sessions, 24-hour regular sessions, CONT17 campaign, and SINEX files spanning 1996–2022. The evaluation results confirm that GASV achieves EOP accuracy comparable to that of VLBI analysis software employed by leading data analysis centers at BKG and USNO.
    
    Version 1.0 of GASV will be released via \url{https://github.com/yaodang/GASV}. This initial release supports full-capability, high-quality single-session VLBI analysis, along with a preliminary implementation of global analysis functionality. A complete global analysis module is planned for the next version. The long-term objective of GASV is to integrate multi-technology processing capabilities, enabling unified analysis of VLBI, GNSS, and SLR data to generate combined solutions.
   
   \begin{acknowledgements}
    This work was supported by the Strategic Priority Research Program of the Chinese Academy of Sciences (XDB1070201).
   \end{acknowledgements}
    
    \bibliographystyle{raa}
    \bibliography{reference.bib}
    
    \begin{appendices}
    \section{Appendix: Global solution theory} \label{app1}
    \setcounter{equation}{0}
        For a single session, the Normal equation system is usually:
        \begin{equation}
            \label{1}
            \begin{bmatrix}
             n_{11} & n_{12} & n_{13} & n_{14}\\
             n_{21} & n_{22} & n_{23} & n_{24}\\
             n_{31} & n_{32} & n_{33} & n_{34}\\
             n_{41} & n_{42} & n_{43} & n_{44}
            \end{bmatrix}\cdot \begin{bmatrix}
            x_{1} \\
            x_{2} \\
            x_{3} \\
            x_{4}
            \end{bmatrix}=\begin{bmatrix}
            b_{1} \\
            b_{2} \\
            b_{3} \\
            b_{4}
            \end{bmatrix}
        \end{equation}
        
        where $n$ denotes coefficient matrix, $x$ represents the parameter vector to be estimated, and $b$ is right-hand side vector.
        
        \subsection{Matrix Sort and Reduce}
        When processing data globally, parameters should be categorized into local parameters and global parameters. Local parameters include clock parameters, zenith wet delay, and tropospheric gradients, while global parameters typically consist of station coordinates and velocities, as well as source coordinates.
        
        If $x_{3}$ represents the global parameters in Equation (\ref{1}), we need to adjust its position to reduce the local parameters (e.g., $x_{1}$, $x_{2}$ and $x_{4}$). Specifically, we first shift the column corresponding to $N_{*3}$ to first column, Equation (\ref{1}) then becomes:
        
        \begin{equation}
             \begin{bmatrix}
             n_{13} & n_{11} & n_{12} & n_{14}\\
             n_{23} & n_{21} & n_{22} & n_{24}\\
             n_{33} & n_{31} & n_{32} & n_{34}\\
             n_{43} & n_{41} & n_{42} & n_{44}
            \end{bmatrix}\cdot \begin{bmatrix}
            x_{3} \\
            x_{1} \\
            x_{2} \\
            x_{4}
            \end{bmatrix}=\begin{bmatrix}
            b_{1} \\
            b_{2} \\
            b_{3} \\
            b_{4}
            \end{bmatrix}   
        \end{equation}
        
        Then, rearrange the matrix rows to ensure symmetry:
        \begin{equation}
            \label{3}
            \begin{bmatrix}
             n_{33} & n_{31} & n_{32} & n_{34}\\
             n_{13} & n_{11} & n_{12} & n_{14}\\
             n_{23} & n_{21} & n_{22} & n_{24}\\
             n_{43} & n_{41} & n_{42} & n_{44}
            \end{bmatrix}\cdot \begin{bmatrix}
            x_{3} \\
            x_{1} \\
            x_{2} \\
            x_{4}
            \end{bmatrix}=\begin{bmatrix}
            b_{3} \\
            b_{1} \\
            b_{2} \\
            b_{4}
            \end{bmatrix}
        \end{equation}
        
        After the column and row shifts, the global and local parameters are separated. Equation (\ref{3}) can then be rewritten as:
        \begin{equation}
            \label{4}
            \begin{bmatrix}
             \mathbf{N}_{11}  & \mathbf{N}_{12}\\
             \mathbf{N}_{21} & \mathbf{N}_{22}
            \end{bmatrix}\cdot \begin{bmatrix}
             \mathbf{x}_{glob}\\
             \mathbf{x}_{local}
            \end{bmatrix}=\begin{bmatrix}
             \mathbf{B}_{1}\\
             \mathbf{B}_{2}
            \end{bmatrix}
        \end{equation}   
        \begin{equation}    
            \mathbf{N}_{11}=[n_{33}],\mathbf{N}_{22}=\begin{bmatrix}
             n_{11} & n_{12} & n_{14}\\
             n_{21} & n_{22} & n_{24}\\
             n_{41} & n_{42} & n_{44}
            \end{bmatrix}
        \end{equation}    
        \begin{equation}
            \mathbf{x}_{glob}=\begin{bmatrix}x_{3}\end{bmatrix},\mathbf{x}_{local}=\begin{bmatrix}
            x_{1}\\
            x_{2}\\
            x_{4}
            \end{bmatrix},
            \mathbf{B}_{1}=\begin{bmatrix}b_{3}\end{bmatrix},\mathbf{B}_{2}=\begin{bmatrix}
            b_{1}\\
            b_{2}\\
            b_{4}
            \end{bmatrix}
        \end{equation}
        
        After reducing  $x_{local}$ , Equation (\ref{4}) becomes:
        
        \begin{align}
            \mathbf{N}_{reduce}\cdot \mathbf{x}_{glob}=\mathbf{B}_{reduce} \label{5}\\
            \mathbf{N}_{reduce}=\mathbf{N}_{11}-\mathbf{N}_{12}\mathbf{N}_{22}^{-1}\mathbf{N}_{21} \label{6}\\
            \mathbf{B}_{reduce}=\mathbf{B}_{1}-\mathbf{N}_{12}\mathbf{N}_{22}^{-1}\mathbf{B}_{2} \label{7}
        \end{align}
        
        \subsection{Parameter Linear}
        If we want to transform the parameter vector $\mathbf{x}_{glob}$ in Equation (\ref{5}) into a new parameter vector $\mathbf{y}$ via a linear relationship, such as:
        \begin{align}
        \label{8}
            \mathbf{x}_{glob}&=\mathbf{C}\mathbf{y}+\mathbf{c}
        \end{align}
        
        Then $\mathbf{N}_{reduce}$ and $\mathbf{B}_{reduce}$ will be:
        \begin{align}
            \mathbf{N}_{new}&=\mathbf{C}^{T}\mathbf{N}_{reduce}\mathbf{C} \\
            \mathbf{B}_{new}&=\mathbf{C}^{T}(\mathbf{B}_{reduce}-\mathbf{N}_{reduce}\cdot \mathbf{c})
        \end{align}
        
        Typically, only station coordinates (excluding velocities) are estimated and written to SINEX file in single-session analysis. Therefore, in the global solution, station coordinates need to be replaced with epoch-specific positions and velocities. According to Equation (\ref{8}), the relationship between the station position at an epoch and its velocity is:
        \begin{equation}
            \mathbf{x}_{glob} =\mathbf{C}\mathbf{y}=\begin{bmatrix}
                 1 &  &  & t_{i}-t_{0} &  &  & \\
                  & 1 &  &  & t_{i}-t_{0} &  & \\
                  &  & 1 &  &  & t_{i}-t_{0} & \\
                  &  &  &  &  &  & ...
                \end{bmatrix}\cdot \begin{bmatrix}
                 \Delta x_{i}\\
                 \Delta y_{i}\\
                 \Delta z_{i}\\
                 \Delta v_{x_{i}}\\
                 \Delta v_{y_{i}}\\
                 \Delta v_{z_{i}}\\
                 ...
            \end{bmatrix}
        \end{equation}
        
        where $i$ is station flag, $t_{i}$ denotes the epoch of the station position estimate within the session, $t_{0}$ represents the reference epoch, and $\Delta x$ and $\Delta v_{x}$ are the differences in position and velocity, respectively:
        \begin{equation}
            \Delta x=x_{e}-x_{a},\Delta v_{x}=v_{ex}-v_{ax}
        \end{equation}
        
        where $x_{e}$ and $\Delta v_{ex}$ are the estimate position and velocity from global solution, $x_{c}$ and $\Delta v_{cx}$ are the apriori position and velocity from the terrestrial reference frame.
         
        \subsection{Stack}
        Consider an existing normal equation system represented by Equation (\ref{13}): 
        \begin{equation}
        \label{13}
            \mathbf{N} = \begin{bmatrix}
             N_{11} &  & ... &  &  & N_{16}\\
              &  &  &  &  & \\
             \vdots &  & \ddots  &  &  & \vdots\\
              &  &  &  &  & \\
              &  &  &  &  & \\
             N_{61} &  & ... &  &  & N_{66}
            \end{bmatrix},\mathbf{x}=\begin{bmatrix}
             x_{1}\\
             x_{2}\\
             x_{3}\\
             x_{4}\\
             x_{5}\\
             x_{6}\end{bmatrix},\mathbf{b}=\begin{bmatrix}
             b_{1}\\
             b_{2}\\
             b_{3}\\
             b_{4}\\
             b_{5}\\
             b_{6}
            \end{bmatrix}
        \end{equation}
        
        Another normal equation system is given by Equation (\ref{14}):
        \begin{equation}
        \label{14}
            \mathbf{N}_{i}=\begin{bmatrix}
             N_{i11} & N_{i12} & N_{i13}\\
             N_{i21} & N_{i22} & N_{i23}\\
             N_{i31} & N_{i32} & N_{i33}
            \end{bmatrix},\mathbf{x}_{i}=\begin{bmatrix}
             x_{1}\\
             x_{3}\\
             x_{6}
            \end{bmatrix},\mathbf{b}_{i}=\begin{bmatrix}
             b_{i1}\\
             b_{i2}\\
             b_{i3}
            \end{bmatrix}
        \end{equation}
        
        Before stacking, first expand Equation (\ref{14}) and then stack it onto Equation (\ref{13}) to form the final normal equation $\mathbf{N}_{all}$ and $\mathbf{b}_{all}$, as shown in Equation (\ref{15}):
        \begin{equation}
        \label{15}
            \mathbf{N}_{all}=\mathbf{N}+\begin{bmatrix}
             N_{i11} & 0 & N_{i12} & 0 & 0 & N_{i13}\\
             0 & 0 & 0 & 0 & 0 & 0\\
             N_{i21} & 0 & N_{i22} & 0 & 0 & N_{i23}\\
             0 & 0 & 0 & 0 & 0 & 0\\
             0 & 0 & 0 & 0 & 0 & 0\\
             N_{i31} & 0 & N_{i32} & 0 & 0 & N_{i33}
            \end{bmatrix},\mathbf{b}_{all} = \mathbf{b}+\begin{bmatrix}
             b_{i1}\\
             0 \\
             b_{i2}\\
             0 \\
             0 \\
             b_{i3}
            \end{bmatrix}
        \end{equation}
        
        \subsection{Constraint}
        When applying no-net-translation (NNT) and no-net-rotation (NNR) to station coordinate, it obey the conditions (\cite{nothnagel_2025}):
        \begin{equation}
            \sum_{i=1}^{N} \mathbf{\Delta x} ^{(i)}=\sum_{i=1}^{N}\begin{pmatrix}
            \Delta x ^{(i)} \\
            \Delta y ^{(i)} \\
            \Delta z ^{(i)}
            \end{pmatrix} =\begin{pmatrix}
            0 \\
            0 \\
            0
            \end{pmatrix}
        \end{equation}
        \begin{equation}
            \sum_{i=1}^{N} (\mathbf{r}^{(i)} \times \mathbf{\Delta x}^{(i)})=\sum_{i=1}^{N}\begin{pmatrix}
            \begin{pmatrix}
            x ^{(i)} \\
            y ^{(i)} \\
            z ^{(i)}
            \end{pmatrix}\times \begin{pmatrix}
            \Delta x ^{(i)} \\
            \Delta y ^{(i)} \\
            \Delta z ^{(i)}
            \end{pmatrix}
            \end{pmatrix}  =\begin{pmatrix}
            0 \\
            0 \\
            0
            \end{pmatrix}
        \end{equation}
        where $\mathbf{\Delta x}$ is the vectorial residuals of the transformation from one frame $\mathbf{x}$ to the other $\mathbf{\tilde{x}}$, and $\mathbf{r}$ is priori coordinte. The final NNR/NNT datum model for the coordinates is:
        \begin{equation}
        \sum_{i=1}^{N} \mathbf{B} _{coord}^{T} \cdot \mathbf{\Delta x}=\mathbf{0}
        \end{equation}
         
        \begin{equation}
            \mathbf{B} _{coord}^{T} = 
            \begin{pmatrix}
            1  & 0 & 0 \\
            0  & 1 & 0 \\
            0  & 0 & 1 \\
            0  & -z^{(i)} & y^{(i)} \\
            z^{(i)}  & 0 & -x^{(i)} \\
            -y^{(i)}  & x^{(i)} & 0
            \end{pmatrix},\mathbf{\Delta x}=\begin{pmatrix}
            \Delta x^{(i)}  \\
            \Delta y^{(i)} \\
            \Delta z^{(i)}
            \end{pmatrix}
        \end{equation}
        
        When applying no-net-translation (NNT) and no-net-rotation (NNR) to station velocities, it obey the conditions:
        \begin{equation}
            \sum_{i=1}^{N} \mathbf{\Delta v} ^{(i)}=\sum_{i=1}^{N}\begin{pmatrix}
            \Delta vx ^{(i)} \\
            \Delta vy ^{(i)} \\
            \Delta vz ^{(i)}
            \end{pmatrix} =\begin{pmatrix}
            0 \\
            0 \\
            0
            \end{pmatrix}
        \end{equation}
        
        \begin{equation}
            \sum_{i=1}^{N} (\mathbf{\Delta x}^{(i)} \times \mathbf{\tilde{v} }^{(i)}+\mathbf{r}^{(i)} \times \mathbf{\Delta v}^{(i)})=\begin{pmatrix}
            \begin{pmatrix}
            \Delta x^{(i)} \\
            \Delta y^{(i)} \\
            \Delta z^{(i)} 
            \end{pmatrix}\times \begin{pmatrix}
            \Delta \tilde{vx} ^{(i)} \\
            \Delta \tilde{vx}^{(i)} \\
            \Delta \tilde{vx}^{(i)} 
            \end{pmatrix}+\begin{pmatrix}
            x^{(i)} \\
            y^{(i)} \\
            z^{(i)} 
            \end{pmatrix}\times \begin{pmatrix}
            \Delta vx ^{(i)} \\
            \Delta vx ^{(i)} \\
            \Delta vx ^{(i)} 
            \end{pmatrix}
            \end{pmatrix}  =\begin{pmatrix}
            0 \\
            0 \\
            0
            \end{pmatrix}
        \end{equation}
        
         where $\mathbf{\Delta v}$ is the vectorial residuals of the transformation from one frame $\mathbf{v}$ to the other $\mathbf{\tilde{v}}$, and $\mathbf{\tilde{v}}$ is priori velocities. The final NNR/NNT datum model for the velocities is:
        \begin{equation}
        \sum_{i=1}^{N} \mathbf{B} _{vel}^{T} \cdot \mathbf{\Delta x}=\mathbf{0} 
        \end{equation}
        \begin{equation}
            \mathbf{B} _{vel}^{T} = 
            \begin{pmatrix}
            0  & 0 & 0 & 1  & 0 & 0 \\
            0  & 0 & 0 & 0  & 1 & 0 \\
            0  & 0 & 0 & 0  & 0 & 1 \\
            0  & \tilde{vz}^{(i)}   & -\tilde{vy}^{(i)} & 0  & -z^{(i)} & y^{(i)} \\
            -\tilde{vz}^{(i)}  & 0 & \tilde{vx}^{(i)} & z^{(i)}  & 0 & -x^{(i)} \\
            \tilde{vy}^{(i)} & -\tilde{vx}^{(i)} & 0 & -y^{(i)}  & x^{(i)} & 0
            \end{pmatrix},\mathbf{\Delta x}=\begin{pmatrix}
            \Delta x^{(i)}  \\
            \Delta y^{(i)} \\
            \Delta z^{(i)} \\
            \Delta vx^{(i)} \\
            \Delta vy^{(i)} \\
            \Delta vz^{(i)}
            \end{pmatrix}
        \end{equation}
        
        When applying no-net-rotation (NNR) to radio sources, it obey the conditions:
        \begin{equation}
            \sum_{i=1}^{N}\mathbf{C}^{T} \times \mathbf{\Delta k}=\mathbf{0}
        \end{equation}
        \begin{equation}
            \mathbf{C}^{T}=\begin{pmatrix}
            -cos\tilde{\alpha}_{i} sin\tilde{\delta}_{i}cos\tilde{\delta}_{i}  &  sin\tilde{\alpha}_{i} \\
            -sin\tilde{\alpha}_{i} sin\tilde{\delta}_{i}cos\tilde{\delta}_{i}  & -cos\tilde{\alpha}_{i}\\
            cos^{2} \tilde{\delta}_{i}  & 0
            \end{pmatrix},\mathbf{\Delta k}=\begin{pmatrix}
            \Delta \alpha _{i}  \\
            \Delta \delta _{i} 
            \end{pmatrix}
        \end{equation}
        
        where $\Delta \alpha=\alpha-\tilde{\alpha}$, $\Delta \delta=\delta-\tilde{\delta}$, $\tilde{\alpha}$ and $\tilde{\delta}$ are the original VLBI positions. Finally, the complete equation is:
        \begin{equation}
            \begin{bmatrix}
             \mathbf{N}_{all} & H\\
             H^{T} & \mathbf{0} 
            \end{bmatrix}\cdot \begin{bmatrix}
             \mathbf{x}\\
            \mathbf{0} 
            \end{bmatrix}=\begin{bmatrix}
            \mathbf{b}_{all} \\
            \mathbf{0} 
            \end{bmatrix} 
        \end{equation}
        \begin{equation}
        \mathbf{H}=(\mathbf{H}_{coord}, \mathbf{H}_{vel},\mathbf{C})
        \end{equation}
        
    \subsection{Formal error}

   SINEX files typically provide the “WEIGHTED SQUARE SUM OF O-C” value, corresponding to:
    
    \begin{equation}
        \ell ^{T}P \ell_{reduce} = (y^{T} {\textstyle \sum_{yy}^{-1}} y)_{reduce}
        \label{sinex_wo}
    \end{equation}
    
    For a single SINEX file, after reducing the local parameters, Equation (\ref{sinex_wo}) becomes:
    
    \begin{equation}
        v ^{T}P v_{reduce} = y^{T} {\textstyle \sum_{yy}^{-1}} y - b_{2}^{T}N_{22}^{-1}b_{2}
    \end{equation}
    
    If the combined global parameters $\tilde{x} _{1}$ are available, the total weighted sum of squares of residuals for all SINEX files is:
    \begin{equation}
    v ^{T}P v = v ^{T}P v_{reduce} - \tilde{x} _{1}^{T}  b_{1}
    \end{equation}
    
    The posteriori variance of unit weight is:
    \begin{equation}
    \sigma _{0}^{2}=\frac{v^{T} P v}{n-u}  
    \end{equation}
    
    The formal error of parameter is then:
    \begin{equation}
    \sigma _{x_{i}}=\sigma _{0}N^{-1}
    \end{equation}

    \end{appendices}

\end{document}